\documentclass[12pt,letterpaper]{article}
\usepackage{epsfig,rotating,setspace,latexsym,amsmath,epsf,amssymb,bm,theorem}
\usepackage{cite,appendix}

\usepackage{amsfonts}

\title{An Alternative Proof for the Capacity Region of the Degraded Gaussian MIMO Broadcast Channel\thanks{This work was supported by
NSF Grants CCF 04-47613, CCF 05-14846, CNS 07-16311 and CCF
07-29127.}}

\author{Ersen Ekrem \qquad Sennur Ulukus \\
\normalsize Department of Electrical and Computer Engineering\\
\normalsize University of Maryland, College Park, MD 20742 \\
\normalsize {\it ersen@umd.edu} \qquad {\it ulukus@umd.edu}}

\newcommand{\brho}{\bm \rho}

\newcommand{\bbsigma}{\bm \Sigma}

\newcommand{\bbi}{{\mathbf{I}}}
\newcommand{\bzero}{{\mathbf{0}}}
\newcommand{\bbv}{{\mathbf{V}}}
\newcommand{\bv}{{\mathbf{v}}}

\newcommand{\bbh}{{\mathbf{H}}}

\newcommand{\bbk}{{\mathbf{K}}}

\newcommand{\bbn}{{\mathbf{N}}}

\newcommand{\bba}{{\mathbf{A}}}
\newcommand{\bbd}{{\mathbf{D}}}

\newcommand{\bbt}{{\mathbf{T}}}

\newcommand{\bbs}{{\mathbf{S}}}
\newcommand{\bu}{{\mathbf{u}}}
\newcommand{\bbj}{{\mathbf{J}}}
\newcommand{\bbu}{{\mathbf{U}}}
\newcommand{\bx}{{\mathbf{x}}}
\newcommand{\bbx}{{\mathbf{X}}}

\newcommand{\bby}{{\mathbf{Y}}}

\newtheorem{Theo}{Theorem}

\newtheorem{Lem}{Lemma}

\newtheorem{Def}{Definition}

\begin{document}


\maketitle

\begin{abstract}
We provide an alternative proof for the capacity region of the
degraded Gaussian multiple-input multiple-output (MIMO) broadcast
channel. Our proof does not use the channel enhancement technique
as opposed to the original proof of Weingertan {\it et. al.} and
the alternative proof of Liu {\it et. al}. Our proof starts with
the single-letter description of the capacity region of the
degraded broadcast channel, and directly evaluates it for the
degraded Gaussian MIMO broadcast channel by using two main
technical tools. The first one is the generalized de Bruijn
identity due to Palomar~\emph{et. al.} which provides a connection
between the differential entropy and the Fisher information
matrix. The second tool we use is an inequality due to Dembo which
lower bounds the differential entropy in terms of the Fisher
information matrix.
\end{abstract}

\newpage
\section{Introduction}

The Gaussian multiple-input multiple-output (MIMO) broadcast
channel consists of one transmitter and an arbitrary number of
receivers, where the transmitter and receivers are equipped with
multiple antennas. In this channel, each link between the
transmitter and each receiver is modeled by a linear additive
Gaussian channel. In general, the Gaussian MIMO broadcast channel
is non-degraded, thus, we do not have a single-letter description
of the capacity region. Despite this lack of a single-letter
description, the capacity region of the Gaussian MIMO broadcast
channel is successfully obtained in~\cite{Shamai_MIMO}.
Subsequently, an alternative proof is given
in~\cite{Liu_Extremal_Inequality}. In both proofs, the {\it
channel enhancement} technique~\cite{Shamai_MIMO} is the main
tool.

Reference~\cite{Shamai_MIMO} obtains the capacity region of the
Gaussian MIMO broadcast channel in three main steps. As the first
step,~\cite{Shamai_MIMO} finds the capacity region of the degraded
Gaussian MIMO broadcast channel. To this end,~\cite{Shamai_MIMO}
first shows that as opposed to the scalar Gaussian broadcast
channel~\cite{Bergmans}, the entropy-power inequality falls short
of providing a converse proof for the degraded Gaussian vector,
i.e., MIMO, broadcast channel. This insufficiency of the
entropy-power inequality is alleviated by the invention of the
{\it channel enhancement} technique~\cite{Shamai_MIMO}. Using this
technique,~\cite{Shamai_MIMO} constructs a new degraded Gaussian
MIMO broadcast channel for each point on the boundary of the
Gaussian rate region\footnote{The Gaussian rate region refers to
the achievable rate region obtained by superposition coding and
successive decoding with Gaussian codebooks.} of the original
degraded channel, where the boundaries of the Gaussian rate
regions of both channels intersect at that point, and the capacity
region of the constructed degraded channel includes the capacity
region of the original one. Then,~\cite{Shamai_MIMO} completes the
first step of the proof by showing that the Gaussian rate region
is the capacity region of the constructed degraded channel, for
which Bergmans' converse~\cite{Bergmans} can be adapted as opposed
to the original channel.


Secondly,~\cite{Shamai_MIMO} considers the aligned Gaussian MIMO
broadcast channel, where the transmitter and all receivers have
the same number of antennas. This channel is not degraded, thus,
there is no single-letter expression for its capacity region.
Reference~\cite{Shamai_MIMO} shows that the achievable rate region
obtained by using dirty paper coding (DPC), i.e., the DPC region,
is the capacity region of the aligned channel. For this
purpose,~\cite{Shamai_MIMO} uses the channel enhancement technique
one more time along with the capacity result obtained for the
degraded Gaussian MIMO broadcast channel in the first step.

Finally,~\cite{Shamai_MIMO} considers the general, not necessarily
degraded or aligned, Gaussian MIMO broadcast channel and shows
that the DPC region again amounts to the capacity region by using
some limiting arguments in conjunction with the capacity result
obtained for the aligned channel.

Similar to the proof of~\cite{Shamai_MIMO}, the alternative proof
in~\cite{Liu_Extremal_Inequality} uses the channel enhancement
technique as well. The alternative proof
in~\cite{Liu_Extremal_Inequality} can be divided into two parts.
In the first part,~\cite{Liu_Extremal_Inequality} considers an
optimization problem which is the maximization of the difference
of two differential entropy terms (see Theorem~1
of~\cite{Liu_Extremal_Inequality}) which cannot be solved by a
stand-alone use of the entropy-power inequality.
Next,~\cite{Liu_Extremal_Inequality} provides two proofs for the
fact that the Gaussian distribution is the maximizer of this
optimization problem. In both proofs provided
in~\cite{Liu_Extremal_Inequality}, the channel enhancement
technique is used. In the second
part,~\cite{Liu_Extremal_Inequality} considers Marton's outer
bound~\cite{Marton} for the general broadcast channel, and
evaluates it for the aligned Gaussian MIMO broadcast channel by
using the optimization problem solved in the first part. This
evaluation yields the capacity region of the two-user aligned
Gaussian MIMO broadcast channel.

We note that though the proof in~\cite{Liu_Extremal_Inequality} is
for the aligned, not necessarily degraded, Gaussian MIMO broadcast
channel, if this proof is adapted to find the capacity region of
the degraded Gaussian MIMO broadcast channel, again the channel
enhancement technique will be needed. In particular, for the
degraded case, the optimization problem solved in the first part
of the proof in~\cite{Liu_Extremal_Inequality} will change
slightly, however the need for channel enhancement will remain. In
fact, the optimization problem needed for the degraded case is a
special case of the original optimization problem solved in the
first part of the proof in~\cite{Liu_Extremal_Inequality}, which
is given in Corollary~4 of~\cite{Liu_Extremal_Inequality}.

Here, we revisit the degraded Gaussian MIMO broadcast channel and
provide an alternative proof for the capacity region of this
degraded channel, without using the channel enhancement technique.
Though channel enhancement is an elegant technique that finds
itself diverse applications, we believe that our proof is more
direct. On the other hand, our proof is limited to the degraded
case and does not seem to be extendable for the general case. In
other words, to obtain the capacity region for the general case
after finding the capacity region for the degraded case through
our proof, one needs to use the channel enhancement
technique~\cite{Shamai_MIMO}.

Our proof starts with the single-letter description of the
capacity region of the degraded broadcast channel and directly
evaluates it for the Gaussian MIMO case by using two main tools.
The first one is the generalized de Bruijn identity due
to~\cite{Palomar_Gradient} that states a connection between the
differential entropy and the Fisher information matrix. The second
tool we use is an inequality due to~\cite{Dembo,Dembo_Cover} that
gives a lower bound for the differential entropy in terms of the
Fisher information matrix.

Finally, our technique used in this alternative proof can be
useful in other vector Gaussian multi-terminal information theory
problems when proving the optimality of Gaussian random vectors.
In fact, we have used a variant of this technique in proving the
secrecy capacity region of the Gaussian MIMO multi-receiver
wiretap channel in~\cite{MIMO_BC_Secrecy}, and the secrecy
capacity region of the Gaussian MIMO degraded compound
multi-receiver wiretap channel in~\cite{Ekrem_Compound}.


\section{Channel Model and Main Result}
The (aligned) degraded $K$-user Gaussian MIMO broadcast channel is
defined by
\begin{align}
\bby_k=\bbx+\bbn_k,\quad k=1,\ldots,K \label{channel}
\end{align}
where $\bbn_k$ is Gaussian with covariance matrix $\bbsigma_k$,
$k=1,\ldots,K$, and the channel input $\bbx$ and outputs
$\{\bby_k\}_{k=1}^K$ satisfy the Markov chain
\begin{align}
\bbx\rightarrow \bby_1 \rightarrow \ldots \rightarrow \bby_K
\label{markov_chain_later}
\end{align}
which is equivalent to the covariance matrices
$\{\bbsigma_k\}_{k=1}^K$ satisfying the following order
\begin{align}
\bzero \prec \bbsigma_1 \preceq \ldots \preceq \bbsigma_K
\end{align}
The channel input is subject to a covariance constraint
\begin{align}
E\left[\bbx \bbx^\top\right] \preceq \bbs
\label{covariance_constraint}
\end{align}
where we assume $\bbs\succ \bzero$. The covariance constraint in
(\ref{covariance_constraint}) is more general than many other
constraints including the trace constraint, in the sense that,
once the capacity region is found for the constraint in
(\ref{covariance_constraint}), capacity regions arising from the
use of other constraints subsumed by (\ref{covariance_constraint})
can be obtained by using this capacity region~\cite{Shamai_MIMO}.

We next note that the definition of degradedness can be
generalized to the case where receivers get arbitrary linear
combinations of the channel inputs, i.e.,
\begin{align}
\bby_k&=\bbh_k \bbx+\bbn_k,\quad k=1,\ldots,K
\label{channel_general}
\end{align}
The broadcast channel defined in (\ref{channel_general}) is said
to be degraded, i.e., satisfies the Markov chain in
(\ref{markov_chain_later}), if there exist matrices
$\{\bbd_k\}_{k=1}^{K-1}$ such that $\bbd_k\bbh_k=\bbh_{k+1}$ and
$\bbd_k\bbd_k^\top\preceq \bbi$~\cite{Liu_Compound}. However, once
the capacity region of the aligned degraded Gaussian MIMO
broadcast channel defined by (\ref{channel}) is obtained, the
capacity region of the general degraded Gaussian MIMO broadcast
channel defined by (\ref{channel_general}) can be obtained by
following the analysis given in Section~5 of~\cite{Liu_Compound},
which essentially relies on some limiting arguments. Since the key
step to obtain the capacity region of the general degraded
Gaussian MIMO broadcast channel defined by~(\ref{channel_general})
is to establish the capacity region of the aligned degraded
Gaussian MIMO broadcast channel defined by (\ref{channel}), here
we consider only the latter channel model.

The capacity region of the Gaussian MIMO broadcast channel is
established in~\cite{Shamai_MIMO} for the most general case. For
the degraded case, it is given as follows.
\begin{Theo}[\!\!\cite{Shamai_MIMO}, Theorem~2]
\label{theorem_arbitrary_K} The capacity region of the $K$-user
degraded Gaussian MIMO broadcast channel is given by the union of
rate tuples $(R_1,\ldots,R_K)$ satisfying
\begin{align}
R_k \leq \frac{1}{2} \log
\frac{|\sum_{i=1}^k\bbk_i+\bbsigma_k|}{|\sum_{i=1}^{k-1}\bbk_i+\bbsigma_k|}
\end{align}
where the union is over all positive semi-definite matrices
$\{\bbk_i\}_{i=1}^K$ such that $\sum_{i=1}^K\bbk_i=\bbs$.
\end{Theo}

In the next section, we provide an alternative proof for this
theorem for $K=2$, and in Section~\ref{sec:arbitrary_K} we extend
this proof to the case $K>2$. In both cases, we directly evaluate
the capacity region of the degraded broadcast channel which is
stated in the following theorem, for the Gaussian MIMO channel at
hand.
\begin{Theo}[\!\!\cite{cover_book},~Theorem~15.6.2]
\label{theo_degraded_bc} The capacity region of the degraded
broadcast channel is given by the union of rate tuples
$(R_1,\ldots,R_K)$ satisfying
\begin{align}
R_k \leq I(U_k;Y_k|U_{k+1}),\quad k=1,\ldots,K
\end{align}
where $U_{K+1}=\phi, U_1=X$, and the union is over all
$\{U_k\}_{k=2}^{K},X$ such that
\begin{align}
U_K \rightarrow \ldots \rightarrow U_2 \rightarrow X \rightarrow
Y_1 \rightarrow \ldots \rightarrow Y_K
\end{align}
\end{Theo}

\section{Proof of Theorem~\ref{theorem_arbitrary_K} for $K=2$}

\subsection{Background}

\label{sec:background} We need some properties of the Fisher
information and the differential entropy, which are provided next.

\begin{Def}[\!\!\cite{MIMO_BC_Secrecy}, Definition~3]
Let $(\bbu,\bbx)$ be an arbitrarily correlated length-$n$ random
vector pair with well-defined densities. The conditional Fisher
information matrix of $\bbx$ given $\bbu$ is defined as
\begin{align}
\bbj(\bbx|\bbu)=E\left[\brho(\bbx|\bbu)\brho(\bbx|\bbu)^\top\right]
\end{align}
where the expectation is over the joint density $f(\bu,\bx)$, and
the conditional score function $\brho(\bx|\bu)$ is
\begin{align}
\brho(\bx|\bu)&=\nabla \log f(\bx|\bu)=\left[~\frac{\partial\log
f(\bx|\bu)}{\partial x_1}~~\ldots~~\frac{\partial\log
f(\bx|\bu)}{\partial x_n}~\right]^\top
\end{align}
\end{Def}

We first present the conditional form of the Cramer-Rao
inequality, which is proved in~\cite{MIMO_BC_Secrecy}.
\begin{Lem}[\!\!\cite{MIMO_BC_Secrecy}, Lemma~13]
\label{lemma_conditional_crb_vector} Let $\bbu,\bbx$ be
arbitrarily correlated random vectors with well-defined densities.
Let the conditional covariance matrix of $\bbx$ be ${\rm
Cov}(\bbx|\bbu)\succ \bzero$, then we have
\begin{align}
\bbj(\bbx|\bbu)\succeq {\rm Cov}(\bbx|\bbu)^{-1}
\end{align}
which is satisfied with equality if $(\bbu,\bbx)$ is jointly
Gaussian with conditional covariance matrix ${\rm
Cov}(\bbx|\bbu)$.
\end{Lem}

The following lemma will be used in the upcoming proof. The
unconditional version of this lemma, i.e., the case $\bbt=\phi$,
is proved in Lemma 6 of~\cite{MIMO_BC_Secrecy}.
\begin{Lem}[\!\!\cite{MIMO_BC_Secrecy},~Lemma~6]
\label{lemma_change_in_fisher} Let $\bbt,\bbu,\bbv_1,\bbv_2$ be
random vectors such that $(\bbt,\bbu)$ and \break$(\bbv_1,\bbv_2)$
are independent. Moreover, let $\bbv_1,\bbv_2$ be Gaussian random
vectors with covariance matrices $\bbsigma_1,\bbsigma_2$ such that
$\bzero \prec \bbsigma_1 \preceq \bbsigma_2$. Then, we have
\begin{align}
\bbj^{-1}(\bbu+\bbv_2|\bbt)-\bbsigma_2 \succeq
\bbj^{-1}(\bbu+\bbv_1|\bbt)-\bbsigma_1
\end{align}
\end{Lem}

The following lemma will also be used in the upcoming proof.
\begin{Lem}[\!\!\cite{MIMO_BC_Secrecy},~Lemma~8]
\label{Shamai_s_lemma} Let $\bbk_1,\bbk_2$ be positive
semi-definite matrices satisfying
$\bzero\preceq\bbk_1\preceq\bbk_2$, and $\mathbf{f}(\bbk)$ be a
matrix-valued function such that $\mathbf{f}(\bbk)\succeq\bzero$
for $\bbk_1\preceq\bbk\preceq \bbk_2$. Then, we have
\begin{align}
\int_{\bbk_1}^{\bbk_2}\mathbf{f}(\bbk)d\bbk \geq 0
\end{align}
\end{Lem}

The following generalization of the de Bruijn identity~\cite{Stam,
Blachman} is due to~\cite{Palomar_Gradient}, where the
unconditional form of this identity, i.e., $\bbu=\phi$, is proved.
Its generalization to this conditional form for an arbitrary
$\bbu$ is rather straightforward, and is given in Lemma~16
of~\cite{MIMO_BC_Secrecy}.
\begin{Lem}[\!\!\cite{MIMO_BC_Secrecy}, Lemma~16]
\label{gradient_fisher_conditional} Let $(\bbu,\bbx)$ be an
arbitrarily correlated random vector pair with finite second order
moments, and also be independent of the random vector $\bbn$ which
is zero-mean Gaussian with covariance matrix
$\bbsigma_N\succ\bzero$. Then, we have
\begin{align}
\nabla_{\bbsigma_N} h(\bbx+\bbn|\bbu)=\frac{1}{2}
\bbj(\bbx+\bbn|\bbu)
\end{align}
\end{Lem}

The following lemma is due to~\cite{Dembo,Dembo_Cover} which lower
bounds the differential entropy in terms of the Fisher information
matrix.
\begin{Lem}[\!\!\cite{Dembo,Dembo_Cover}]
\label{lemma_dembo} Let $(U,\bbx)$ be an $(n+1)$-dimensional
random vector, where the conditional Fisher information matrix of
$\bbx$, conditioned on $U$, exists. Then, we have
\begin{align}
h(\bbx|U) \geq \frac{1}{2} \log (2\pi e)^n |\bbj^{-1}(\bbx|U)|
\end{align}
\end{Lem}
In~\cite{Dembo,Dembo_Cover}, the unconditional version of this
lemma, i.e., $U=\phi$, is provided. A proof for its generalization
to this conditional form is given in
Appendix~\ref{proof_of_dembos_lemma}.

\subsection{Proof for $K=2$}
\label{sec:proof_two_user}

We first rewrite the capacity region of the degraded broadcast
channel given in Theorem~\ref{theo_degraded_bc} for two users as a
union of rate pairs $(R_1,R_2)$ satisfying
\begin{align}
R_1 & \leq I(X;Y_1|U) \label{R1_discrete}\\
R_2 & \leq I(U;Y_2) \label{R2_discrete}
\end{align}
where we dropped the subscript of the auxiliary random variable
$U_2$ and denoted it simply as $U$. The involved random variables
satisfy the Markov chain $U\rightarrow X \rightarrow Y_1
\rightarrow Y_2$. To obtain the capacity region of the degraded
Gaussian MIMO broadcast channel, we need to evaluate this region.
In particular, we will show that the optimal random vector
$(U,\bbx)$ that exhausts this region is Gaussian, and the
corresponding capacity region is given by the union of rate pairs
$(R_1,R_2)$ satisfying
\begin{align}
R_1 \leq \frac{1}{2} \log \frac{|\bbk+\bbsigma_1|}{|\bbsigma_1|}
\label{R1_two_user}
\\
R_2 \leq \frac{1}{2} \log
\frac{|\bbs+\bbsigma_2|}{|\bbk+\bbsigma_2|} \label{R2_two_user}
\end{align}
where the union is over all $\bbk$ such that $\bzero \preceq \bbk
\preceq \bbs$. We note that the region described by
(\ref{R1_two_user})-(\ref{R2_two_user}) comes from
Theorem~\ref{theorem_arbitrary_K} by dropping the subscript of
$\bbk_1$ and denoting it simply as $\bbk$.

We begin with the bound on $R_2$. Starting from
(\ref{R2_discrete}), we get
\begin{align}
R_2 &\leq I(U;\bby_2)\\
& =h(\bby_2)-h(\bby_2|U) \\
&\leq \frac{1}{2} \log (2\pi e)^n |\bbs+\bbsigma_2|-h(\bby_2|U)
\label{max_entropy_implies}
\end{align}
where the inequality in~(\ref{max_entropy_implies}) comes from the
maximum entropy theorem~\cite{cover_book}. We now bound
$h(\bby_2|U)$ in (\ref{max_entropy_implies}). We first get an
upper bound as
\begin{align}
h(\bby_2|U)\leq h(\bby_2) \leq \frac{1}{2} \log (2\pi e)^n
|\bbs+\bbsigma_2| \label{upper_bound}
\end{align}
where the first inequality comes from the fact that conditioning
cannot increase entropy, and the second inequality is due to the
maximum entropy theorem~\cite{cover_book}. Furthermore, using
Lemma~\ref{lemma_dembo}, we can get the following lower bound for
$h(\bby_2|U)$
\begin{align}
h(\bby_2|U) \geq \frac{1}{2} \log (2\pi e)^n
|\bbj^{-1}(\bbx+\bbn_2|U)| \label{lower_bound}
\end{align}

We next define the following function
\begin{align}
r(t)=\frac{1}{2}\log (2\pi e)^n |\bba(t)+\bbsigma_2|,\quad 0\leq
t\leq 1 \label{r_t}
\end{align}
where $\bba(t)$ is given as
\begin{align}
\bba(t)=(1-t)
\left[\bbj^{-1}(\bbx+\bbn_2|U)-\bbsigma_2\right]+t\bbs
\end{align}
We first note that
\begin{align}
\bbj^{-1}(\bbx+\bbn_2|U)-\bbsigma_2 & \preceq
\textrm{Cov}(\bbx+\bbn_2|U)-\bbsigma_2 \label{crb_implies} \\
& =
\textrm{Cov}(\bbx|U) \\
& \preceq \textrm{Cov}(\bbx) \label{conditioning_cannot_1}\\
& \preceq \bbs
\end{align}
where (\ref{crb_implies}) is a consequence of
Lemma~\ref{lemma_conditional_crb_vector}, and
(\ref{conditioning_cannot_1}) comes from the fact that the
conditional covariance matrix is smaller than the unconditional
one in the positive semi-definite ordering sense. This implies
that for any $0\leq t\leq 1$, $\bba(t)$ satisfies
\begin{align}
\bbj^{-1}(\bbx+\bbn_2|U)-\bbsigma_2 \preceq \bba(t) \preceq \bbs
\label{the_order}
\end{align}

Using $r(t)$, bounds in (\ref{upper_bound}) and
(\ref{lower_bound}) can be rewritten as
\begin{align}
r(0)\leq h(\bby_2|U) \leq r(1)
\end{align}
Since $r(t)$ is continuous in $t$~\cite{MIMO_BC_Secrecy}, due to
the intermediate value theorem, there exists a $t^*$ such that
\begin{align}
r(t^*)=h(\bby_2|U)=\frac{1}{2}\log (2\pi e)^n
|\bba(t^*)+\bbsigma_2| \label{fixed_point}
\end{align}
where $\bba(t^*)$ satisfies (\ref{the_order}). Plugging
(\ref{fixed_point}) into (\ref{max_entropy_implies}) yields
\begin{align}
R_2 \leq \frac{1}{2} \log
\frac{|\bbs+\bbsigma_2|}{|\bba(t^*)+\bbsigma_2|}
\end{align}
which is the desired bound on $R_2$ given in~(\ref{R2_two_user}).

We now obtain the desired bound on $R_1$. To this end, using
(\ref{the_order}) and Lemma~\ref{lemma_change_in_fisher}, we get
\begin{align}
\bba(t^*) &\succeq \bbj^{-1}(\bbx+\bbn_2|U)-\bbsigma_2 \\
&\succeq \bbj^{-1}(\bbx+\bbn|U)-\bbsigma_N
\label{the_order_fisher}
\end{align}
for any Gaussian random vector $\bbn$ with covariance matrix
$\bbsigma_N$ where $\bbsigma_N \preceq \bbsigma_2$. The order in
(\ref{the_order_fisher}) is equivalent to
\begin{align}
(\bba(t^*)+\bbsigma_N)^{-1} &\preceq \bbj(\bbx+\bbn|U)
\label{the_order_Fisher_1}
\end{align}
Next, starting from~(\ref{R1_discrete}), we get
\begin{align}
R_1&\leq I(\bbx;\bby_1|U)\\
&= h(\bby_1|U)-\frac{1}{2} \log (2\pi e)^n |\bbsigma_1| \\
&= h(\bby_1|U)-h(\bby_2|U)+h(\bby_2|U)-\frac{1}{2} \log (2\pi e)^n |\bbsigma_1| \\
&=h(\bby_1|U)-h(\bby_2|U)+\frac{1}{2}\log |\bba(t^*)+\bbsigma_2|
-\frac{1}{2} \log |\bbsigma_1|
\label{fixed_point_implies} \\
&=-\frac{1}{2}
\int_{\bbsigma_1}^{\bbsigma_2}\bbj(\bbx+\bbn|U)d\bbsigma_N+\frac{1}{2}\log
 \frac{|\bba(t^*)+\bbsigma_2|}{|\bbsigma_1|} \label{de_bruin_implies}\\
&\leq -\frac{1}{2}
\int_{\bbsigma_1}^{\bbsigma_2}(\bba(t^*)+\bbsigma_N)^{-1}d\bbsigma_N+\frac{1}{2}\log
\frac{ |\bba(t^*)+\bbsigma_2|}{|\bbsigma_1|}\label{bound_integral_1} \\
&=\frac{1}{2}\log\frac{
|\bba(t^*)+\bbsigma_1|}{|\bba(t^*)+\bbsigma_2|}+\frac{1}{2}\log\frac{
|\bba(t^*)+\bbsigma_2|}{|\bbsigma_1|}\\
&=\frac{1}{2}\log\frac{
|\bba(t^*)+\bbsigma_1|}{|\bbsigma_1|} \label{desired_on_R1}
\end{align}
where (\ref{fixed_point_implies}) is due to (\ref{fixed_point}),
(\ref{de_bruin_implies}) is obtained by using
Lemma~\ref{gradient_fisher_conditional}, and
(\ref{bound_integral_1}) is due to (\ref{the_order_Fisher_1}) and
Lemma~\ref{Shamai_s_lemma}. Since (\ref{desired_on_R1}) is the
desired bound on $R_1$ given in~(\ref{R1_two_user}), this
completes the proof.

\section{Extension to the $K$-user Case}
\label{sec:arbitrary_K}

We now extend our alternative proof presented in the previous
section to the case $K>2$. For that purpose, we need the following
lemma due to~\cite{MIMO_BC_Secrecy} in addition to the tools
introduced in Section~\ref{sec:background}.
\begin{Lem} [\!\!\cite{MIMO_BC_Secrecy}, Lemma~17]
\label{lemma_cond_increases_fi} Let $(\bbv,\bbu,\bbx)$ be
length-$n$ random vectors with well-defined densities. Moreover,
assume that the partial derivatives of $f(\bu|\bx,\bv)$ with
respect to $x_i,~i=1,\ldots,n$ exist and satisfy
\begin{align}
\max_{1\leq i\leq n}\left|\frac{\partial f(\bu|\bv,\bx)}{\partial
x_i}\right| \leq g(\bu)
\end{align}
for some integrable function $g(\bu)$. Then, if $(\bbv,\bbu,\bbx)$
satisfy the Markov chain $\bbv\rightarrow \bbu \rightarrow \bbx$,
we have
\begin{align}
\bbj(\bbx|\bbu) \succeq \bbj(\bbx|\bbv)
\end{align}
\end{Lem}

First, following the proof in Section~\ref{sec:proof_two_user}, we
can show the existence of a covariance matrix $\bba_K$ such that
\begin{align}
\bbj^{-1}(\bbx+\bbn_K|&U_{K})-\bbsigma_K \preceq  \bba_K \preceq \bbs \label{cov_mat_cond_1} \\
h(\bby_K|U_{K})&=\frac{1}{2}\log (2\pi e)^n |\bba_K+\bbsigma_K| \label{cov_mat_cond_2} \\
h(\bby_{K-1}|U_{K})&\leq \frac{1}{2}\log (2\pi e)^n
|\bba_K+\bbsigma_{K-1}| \label{cov_mat_cond_3}
\end{align}
Since we have
\begin{align}
h(\bby_K) \leq \frac{1}{2} \log (2\pi e)^n |\bbs+\bbsigma_K|
\end{align}
from the maximum entropy theorem~\cite{cover_book}, we can get the
desired bound on $R_K$ as follows
\begin{align}
R_K &\leq I(U_K;\bby_K)\leq \frac{1}{2} \log
\frac{|\bbs+\bbsigma_K|}{|\bba_K+\bbsigma_K|}
\end{align}

When $K=2$ as in Section~\ref{sec:proof_two_user}, showing the
existence of an $\bba_K$ having the properties listed in
(\ref{cov_mat_cond_1})-(\ref{cov_mat_cond_3}) is sufficient to
conclude the proof. However, when $K>2$, we need an additional
tool, which is Lemma~\ref{lemma_cond_increases_fi}, and using this
tool we need to repeat this step until we are left with showing
the desired bound on the first user's rate $R_1$. We now present
the basic step that needs to be repeated. In particular, we now
show the existence of a covariance matrix $\bba_{K-1}$ such that
\begin{align}
\bbj^{-1}(\bbx+\bbn_{K-1}|&U_{K-1})-\bbsigma_{K-1} \preceq  \bba_{K-1} \preceq \bba_K \label{cov_mat_cond_4} \\
h(\bby_{K-1}|U_{K-1})&=\frac{1}{2}\log (2\pi e)^n |\bba_{K-1}+\bbsigma_{K-1}| \label{cov_mat_cond_5} \\
h(\bby_{K-2}|U_{K-1})&\leq \frac{1}{2}\log (2\pi e)^n
|\bba_{K-1}+\bbsigma_{K-2}| \label{cov_mat_cond_6}
\end{align}

We first note the following order
\begin{align}
\bba_K &\succeq \bbj^{-1}(\bbx+\bbn_{K}|U_{K})-\bbsigma_{K} \label{cov_mat_cond_1_implies}\\
&\succeq \bbj^{-1}(\bbx+\bbn_{K-1}|U_{K})-\bbsigma_{K-1}
\label{lemma_change_in_fisher_implies} \\
&\succeq \bbj^{-1}(\bbx+\bbn_{K-1}|U_{K-1})-\bbsigma_{K-1}
\label{lemma_cond_increases_fi_implies}
\end{align}
where (\ref{cov_mat_cond_1_implies}) is due to
(\ref{cov_mat_cond_1}), (\ref{lemma_change_in_fisher_implies})
comes from Lemma~\ref{lemma_change_in_fisher}, and
(\ref{lemma_cond_increases_fi_implies}) follows from
Lemma~\ref{lemma_cond_increases_fi} as we can get
\begin{align}
\bbj(\bbx+\bbn_{K-1}|U_{K-1}) \succeq \bbj(\bbx+\bbn_{K-1}|U_{K})
\end{align}
by noting the Markov chain $U_{K}\rightarrow U_{K-1} \rightarrow
\bbx+\bbn_{K-1}$.

Next, we consider the following lower bound on
$h(\bby_{K-1}|U_{K-1})$ which is due to Lemma~\ref{lemma_dembo}
\begin{align}
\frac{1}{2} \log (2\pi e)^n |\bbj^{-1}&(\bbx+\bbn_{K-1}|U_{K-1})|
\leq h(\bby_{K-1}|U_{K-1}) \label{dembo_is_my_man}
\end{align}
Moreover, we can get the following upper bound
\begin{align}
 h(\bby_{K-1}|U_{K-1}) &= h(\bby_{K-1}|U_{K-1},U_K) \label{markov_chain} \\
&\leq  h(\bby_{K-1}|U_K) \label{conditioning_cannot_2} \\
&\leq \frac{1}{2}\log (2\pi e)^n |\bba_K+\bbsigma_{K-1}|
\label{cov_mat_cond_3_implies}
\end{align}
where (\ref{markov_chain}) is due to the Markov chain
$U_K\rightarrow U_{K-1}\rightarrow \bby_{K-1}$,
(\ref{conditioning_cannot_2}) comes from the fact that
conditioning cannot increase entropy, and
(\ref{cov_mat_cond_3_implies}) is due to (\ref{cov_mat_cond_3}).

We now define the following function
\begin{align}
\hspace{-0.15cm}r_{K-1}(t)=\frac{1}{2} \log (2\pi e)^n
|\tilde{\bba}_{K-1}(t)+\bbsigma_{K-1}|,~0\leq t\leq 1
\end{align}
where $\tilde{\bba}_{K-1}(t)$ is given by
\begin{align}
\tilde{\bba}_{K-1}(t)&=(1-t)\left[\bbj^{-1}(\bbx+\bbn_{K-1}|U_{K-1})-\bbsigma_{K-1}\right]+t
\bba_{K} \label{last_moments_of_christ}
\end{align}
Using $r_{K-1}(t)$, we can recast bounds on
$h(\bby_{K-1}|U_{K-1})$ in (\ref{dembo_is_my_man}) and
(\ref{cov_mat_cond_3_implies}) as
\begin{align}
r_{K-1}(0) \leq h(\bby_{K-1}|U_{K-1}) \leq r_{K-1}(1)
\end{align}
Since $r_{K-1}(t)$ is continuous in $t$~\cite{MIMO_BC_Secrecy},
due to the intermediate value theorem, there exists a $t^{*}$ such
that $h(\bby_{K-1}|U_{K-1})= r_{K-1}(t^*)$, i.e.,
\begin{align}
h(\bby_{K-1}|U_{K-1}) &=\frac{1}{2} \log (2\pi e)^n
|\bba_{K-1}+\bbsigma_{K-1}|
\end{align}
where we define $ \bba_{K-1}=\tilde{\bba}_{K-1}(t^*)$. Thus, we
established (\ref{cov_mat_cond_5}). Furthermore, it is clear that
we also have (\ref{cov_mat_cond_4}) because of
(\ref{lemma_cond_increases_fi_implies}),
(\ref{last_moments_of_christ}) and $0\leq t^*\leq 1$.

We now show (\ref{cov_mat_cond_6}). To this end, we note the
following order
\begin{align}
\bba_{K-1}& \succeq
\bbj^{-1}(\bbx+\bbn_{K-1}|U_{K-1})-\bbsigma_{K-1} \label{cov_mat_cond_4_implies} \\
& \succeq \bbj^{-1}(\bbx+\bbn|U_{K-1})-\bbsigma_N
\label{lemma_change_in_fisher_implies_1}
\end{align}
for any Gaussian $\bbn$ with covariance matrix $\bbsigma_N\preceq
\bbsigma_{K-1}$ due to Lemma~\ref{lemma_change_in_fisher}. The
order in (\ref{lemma_change_in_fisher_implies_1}) is equivalent to
\begin{align}
\bbj(\bbx+\bbn|U_{K-1}) \succeq (\bba_{K-1}+\bbsigma_N)^{-1},~~
\bbsigma_N \preceq \bbsigma_{K-1} \label{the_order_is_an_order}
\end{align}
We now consider $h(\bby_{K-2}|U_{K-1})$ as follows
\begin{align}
h(\bby_{K-2}|U_{K-1})&=h(\bby_{K-2}|U_{K-1})-h(\bby_{K-1}|U_{K-1})
+h(\bby_{K-1}|U_{K-1})
\\
&=-\frac{1}{2}
\int_{\bbsigma_{K-2}}^{\bbsigma_{K-1}}\bbj(\bbx+\bbn|U)d\bbsigma_{N}
+\frac{1}{2} \log (2\pi e)^n |\bba_{K-1}+\bbsigma_{K-1}|
\label{cov_mat_cond_5_implies} \\
&\leq -\frac{1}{2}
\int_{\bbsigma_{K-2}}^{\bbsigma_{K-1}}(\bba_{K-1}+\bbsigma_N)^{-1}d\bbsigma_{N}
+\frac{1}{2} \log (2\pi e)^n |\bba_{K-1}+\bbsigma_{K-1}|
\label{Shamai_s_lemma_implies} \\
&=\frac{1}{2} \log (2\pi e)^n |\bba_{K-1}+\bbsigma_{K-2}|
\label{i_am_sam}
\end{align}
where (\ref{cov_mat_cond_5_implies}) comes from
Lemma~\ref{gradient_fisher_conditional} and
(\ref{cov_mat_cond_5}), and (\ref{Shamai_s_lemma_implies}) is due
to (\ref{the_order_is_an_order}) and Lemma~\ref{Shamai_s_lemma}.
Thus, we showed (\ref{cov_mat_cond_6}) as well. Also, we can
establish the desired bound $R_{K-1}$ as follows
\begin{align}
R_{K-1} &\leq I(U_{K-1};\bby_{K-1}|U_K)\\
&=h(\bby_{K-1}|U_K)-h(\bby_{K-1}|U_{K-1})\label{markov_chain_1}\\
&\leq \frac{1}{2} \log (2\pi
e)^n|\bba_{K}+\bbsigma_{K-1}|-h(\bby_{K-1}|U_{K-1})\label{smthing_sweet}\\
&=\frac{1}{2} \log
\frac{|\bba_{K}+\bbsigma_{K-1}|}{|\bba_{K-1}+\bbsigma_{K-1}|}
\label{smthing_cute}
\end{align}
where (\ref{markov_chain_1}) comes from the Markov chain
$U_{K}\rightarrow U_{K-1} \rightarrow Y_{K-1}$,
(\ref{smthing_sweet}) comes from (\ref{cov_mat_cond_3}), and
(\ref{smthing_cute}) is due to (\ref{cov_mat_cond_5}).

As of now, we outlined the basic step that needs to be repeated
until we are left with getting the desired bound on the first
user's rate $R_1$. Following the analysis from
(\ref{cov_mat_cond_4}) to (\ref{i_am_sam}), we can show the
existence of covariance matrices $\bba_{k},$ for $k=2,\ldots,K,$
such that
\begin{align}
\bbj^{-1}(\bbx+\bbn_{k}|U_k&)-\bbsigma_k\preceq \bba_k \preceq
\bba_{k+1}\label{cov_mat_cond_7}\\
h(\bby_{k}|U_k)&=\frac{1}{2} \log (2\pi e)^n
|\bba_k+\bbsigma_k|\label{cov_mat_cond_8}\\
h(\bby_{k-1}|U_k)&\leq \frac{1}{2} \log (2\pi e)^n
|\bba_k+\bbsigma_{k-1}| \label{cov_mat_cond_9}
\end{align}
where we set $\bba_{K+1}=\bbs.$ Using these relations, we can get
the bound on $R_k$ for any $k=2,\ldots,K,$ as
\begin{align}
R_k&\leq I(U_{k};\bby_{k}|U_{k+1})\\
&=h(\bby_{k}|U_{k+1})-h(\bby_{k}|U_{k})\label{markov_chain_2}\\
&\leq \frac{1}{2}\log (2\pi
e)^n|\bba_{k+1}+\bbsigma_k|-h(\bby_{k}|U_{k})\label{american_beauty}\\
&= \frac{1}{2}\log
\frac{|\bba_{k+1}+\bbsigma_k|}{|\bba_k+\bbsigma_k|}
\label{winterlight}
\end{align}
where we set $U_{K+1}=\phi$. The equality in
(\ref{markov_chain_2}) comes from the Markov chain
$U_{k+1}\rightarrow U_k \rightarrow \bby_k$,
(\ref{american_beauty}) is obtained by using
(\ref{cov_mat_cond_9}), and (\ref{winterlight}) is due to
(\ref{cov_mat_cond_8}). For $k=1$, we can get the bound on $R_1$
as
\begin{align}
R_1 &\leq I(\bbx;\bby_1|U_2)\\
&=h(\bby_1|U_2)-\frac{1}{2} \log (2\pi e)^n |\bbsigma_1|\\
&\leq \frac{1}{2}\log
\frac{|\bba_2+\bbsigma_1|}{|\bbsigma_1|}\label{frank_sinatra}
\end{align}
where (\ref{frank_sinatra}) comes from (\ref{cov_mat_cond_9}).
Finally, we define $\bbk_{k}=\bba_{k+1}-\bba_{k}$ for $k=1,\ldots,
K$ where $\bba_1=\phi,\bba_{K+1}=\bbs$, and plug these into
(\ref{winterlight})-(\ref{frank_sinatra}) which yields the
expressions in Theorem~\ref{theorem_arbitrary_K}.

\section{Conclusions}
We provide an alternative proof for the capacity region of the
degraded Gaussian MIMO broadcast channel. As opposed to the
existing proofs in~\cite{Shamai_MIMO,Liu_Extremal_Inequality}, our
proof does not use the channel enhancement
technique~\cite{Shamai_MIMO}. Our proof starts with the
single-letter description of the degraded broadcast channel's
capacity region and directly evaluates it for the degraded
Gaussian MIMO case. This evaluation is carried out by means of two
main technical tools. The first one is the generalized de Bruijn
identity that gives a connection between the differential entropy
and the Fisher information~\cite{Palomar_Gradient}. The second one
is an inequality due to~\cite{Dembo,Dembo_Cover} that lower bounds
the differential entropy in terms of the Fisher information
matrix.

\appendix
\section{Proof of Lemma~\ref{lemma_dembo}}
\label{proof_of_dembos_lemma}

We define the function $f(\epsilon)$ as follows
\begin{align}
f(\epsilon)=h(\bbx+\sqrt{\epsilon}\bbn|U)-\frac{1}{2}\log\left|(2\pi
e)\left(\bbj^{-1}(\bbx|U)+\epsilon \bbsigma\right)\right|,\quad
\epsilon \geq 0
\end{align}
We need to prove that $f(0)\geq 0$. We will show that
$f(\epsilon)$ is monotonically decreasing in $\epsilon$, and that
$\lim_{\epsilon\rightarrow \infty}f(\epsilon)=0$. This will prove
$f(0)\geq 0$. To this end, we introduce the following lemma which
will be used subsequently.
\begin{Lem}[\!\!\cite{MIMO_BC_Secrecy}, Corollary~4]
\label{lemma_conditional_fii} Let $\bbx,\bby,\bbu$ be length-$n$
random vectors and let the density for any combination of these
random vectors exist. Moreover, let $\bbx$ and $\bby$ be
conditionally independent given $\bbu$. Then, we have
\begin{align}
\bbj(\bbx+\bby|\bbu)\preceq
\left[\bbj(\bbx|\bbu)^{-1}+\bbj(\bby|\bbu)^{-1}\right]^{-1}
\end{align}
\end{Lem}

Fix $\epsilon_1,\epsilon_2$ such that $0<\epsilon_1 \leq
\epsilon_2$. Using Lemma~\ref{gradient_fisher_conditional}, we
have
\begin{align}
h(\bbx+\sqrt{\epsilon_2}\bbn|U)-h(\bbx+\sqrt{\epsilon_1}\bbn|U)=\frac{1}{2}
\int_{\epsilon_1\bbsigma}^{\epsilon_2 \bbsigma}
\bbj(\bbx+\bbt|U)d\bbsigma_T \label{a_integral}
\end{align}
where $\bbt$ is a Gaussian random vector with covariance matrix
$\bbsigma_T$ such that $\epsilon_1 \bbsigma\preceq
\bbsigma_T\preceq \epsilon_2 \bbsigma$, and independent of
$(U,\bbx)$. Using Lemma~\ref{lemma_conditional_fii} in conjunction
with Lemma~\ref{lemma_conditional_crb_vector}, we get
\begin{align}
\bbj(\bbx+\bbt|U)\preceq
\left[\bbj^{-1}(\bbx|U)+\bbsigma_T\right]^{-1} \label{a_fii}
\end{align}
Plugging (\ref{a_fii}) into (\ref{a_integral}) and invoking
Lemma~\ref{Shamai_s_lemma}, we get
\begin{align}
h(\bbx+\sqrt{\epsilon_2}\bbn|U)-h(\bbx+\sqrt{\epsilon_1}\bbn|U)\leq
\frac{1}{2} \log\frac{\left|(2\pi
e)\left(\bbj^{-1}(\bbx|U)+\epsilon_2\bbsigma\right)\right|}{\left|(2\pi
e)\left(\bbj^{-1}(\bbx|U)+\epsilon_1\bbsigma\right)\right|}
\label{a_bounded_integral}
\end{align}
Rearranging (\ref{a_bounded_integral}) yields
\begin{align}
f(\epsilon_2)\leq f(\epsilon_1),\quad \epsilon_1\leq \epsilon_2
\end{align}
which proves that $f(\epsilon)$ is monotonically decreasing in
$\epsilon$.

We now consider upper and lower bounds on $f(\epsilon)$. We have
the following upper bound on $f(\epsilon)$
\begin{align}
f(\epsilon)&=h(\bbx+\sqrt{\epsilon}\bbn|U)-\frac{1}{2}
\log\left|(2\pi
e)\left(\bbj^{-1}(\bbx|U)+\epsilon\bbsigma\right)\right| \\
&\leq \frac{1}{2} \log\frac{|\bbk+\epsilon\bbsigma|}{|\bbj^{-1}(\bbx|U)+\epsilon\bbsigma|}\label{a_max_entropy} \\
&= \frac{1}{2} \log\frac{
|\bbsigma^{-1/2}\bbk\bbsigma^{-1/2}+\epsilon\bbi|}{|\bbsigma^{-1/2}\bbj^{-1}(\bbx|U)\bbsigma^{-1/2}+\epsilon\bbi|}
\\
&=\frac{1}{2} \log \prod_{i=1}^n
\frac{\tilde{\lambda}_i+\epsilon}{\lambda_i+\epsilon}
\label{a_eigenvalues}
\end{align}
where (\ref{a_max_entropy}) comes from the maximum entropy
theorem~\cite{cover_book} and $\bbk$ denotes the covariance matrix
of $\bbx$. In~(\ref{a_eigenvalues}), we denote the eigenvalues of
$\bbsigma^{-1/2}\bbk\bbsigma^{-1/2}$ with
$\{\tilde{\lambda}_i\}_{i=1}^n$, and of
$\bbsigma^{-1/2}\bbj^{-1}(\bbx|U)\bbsigma^{-1/2}$ with
$\{\lambda_i\}_{i=1}^n$. Furthermore, we have the following lower
bound on $f(\epsilon)$
\begin{align}
f(\epsilon)&= h(\bbx+\sqrt{\epsilon}\bbn|U)-\frac{1}{2}
\log\left|(2\pi e)\left(\bbj^{-1}(\bbx|U)+\epsilon\bbsigma\right)\right| \\
&\geq \frac{1}{2} \log\frac{|\epsilon \bbsigma|}{|\bbj^{-1}(\bbx|U)+\epsilon\bbsigma|}\label{conditioning_cannot}\\
&= \frac{1}{2} \log\frac{\epsilon^n
}{|\bbsigma^{-1/2}\bbj^{-1}(\bbx|U)\bbsigma^{-1/2}+\epsilon\bbi|}
\\
&= \frac{1}{2} \log \prod_{i=1}^n\frac{\epsilon
}{\lambda_i+\epsilon} \label{a_eigenvalues_1}
\end{align}
where (\ref{conditioning_cannot}) comes from the fact that
conditioning cannot increase entropy, and in
(\ref{a_eigenvalues_1}), we denote the eigenvalues of
$\bbsigma^{-1/2}\bbj^{-1}(\bbx|U)\bbsigma^{-1/2}$ with
$\{\lambda_i\}_{i=1}^n$. Comparison of (\ref{a_eigenvalues}) and
(\ref{a_eigenvalues_1}) yields
\begin{align}
\frac{1}{2} \log \prod_{i=1}^n\frac{\epsilon }{\lambda_i+\epsilon}
\leq f(\epsilon) \leq \frac{1}{2} \log
\prod_{i=1}^n\frac{\tilde{\lambda}_i+\epsilon
}{\lambda_i+\epsilon}
\end{align}
Taking the limit as $\epsilon\rightarrow \infty$ yields
$\lim_{\epsilon\rightarrow \infty}f(\epsilon)=0$. Combining this
with the fact that $f(\epsilon)$ decreases monotonically in
$\epsilon$ yields $f(0)\geq0$, and consequently,
\begin{align}
h(\bbx|U) \geq \frac{1}{2} \log (2\pi e)^n |\bbj^{-1}(\bbx|U)|
\end{align}
completing the proof.

\bibliographystyle{unsrt}
\bibliography{IEEEabrv,references2}
\end{document}